\documentclass{iau}
\usepackage{graphicx}

\title[AIC process in White Dwarfs] 
{Deriving fundamental parameters of millisecond pulsars via AIC in white dwarfs}

\author[Ali Taani et al.]   
{A. Taani$^1$$^\dag$, Y.H. Zhao$^1$ \and A. Moraghan$^2$}

\affiliation{$^1$National Astronomical Observatories, Chinese Academy of Sciences, Beijing 100012, China
\break $^\dag$email: alitaani@bao.ac.cn\\[\affilskip]
$^2$Center for Galaxy Evolution Research and Department of Astronomy,
Yonsei University, Seoul 120-749, Republic of Korea}

\pubyear{2012}
\volume{290}  
\pagerange{119--126}
\jname{Feeding compact objects: Accretion on all scales}
\editors{C.M. Zhang, T. Belloni, M. Mendez,
and S.N Zhang, eds.}

\begin{document}

\maketitle

\begin{abstract}
We present a study of the observational properties of Millisecond
Pulsars (MSPs) by way of their magnetic fields, spin periods and
masses. These measurements are derived through the scenario of
Accretion Induced Collapse (AIC) of white dwarfs (WDs) in stellar
binary systems, in order to provide a greater understanding of the
characteristics of MSP populations. 
In addition, we demonstrate a
strong evolutionary connection between neutron stars and WDs with
binary companions from a stellar binary evolution perspective via
the AIC process.



\end{abstract}




\section{Introduction}

Observable parameters of binary Millisecond Pulsars (MSPs), e.g.
mass of the pulsar, mass of the companion, spin period, orbital
period, eccentricity, etc., are used to probe the past accretion
history of the MSPs. The purpose of this proceeding is to
demonstrate how to infer some of the observable quantities (spin
period, magnetic field and mass) during the Accretion Induced
Collapse (AIC) of a white dwarf (WD) on its way to become a member
of the MSP family.

\begin{figure*}
\begin{center}
\hspace*{\fill}
\includegraphics[angle=0, width=4.5 cm]{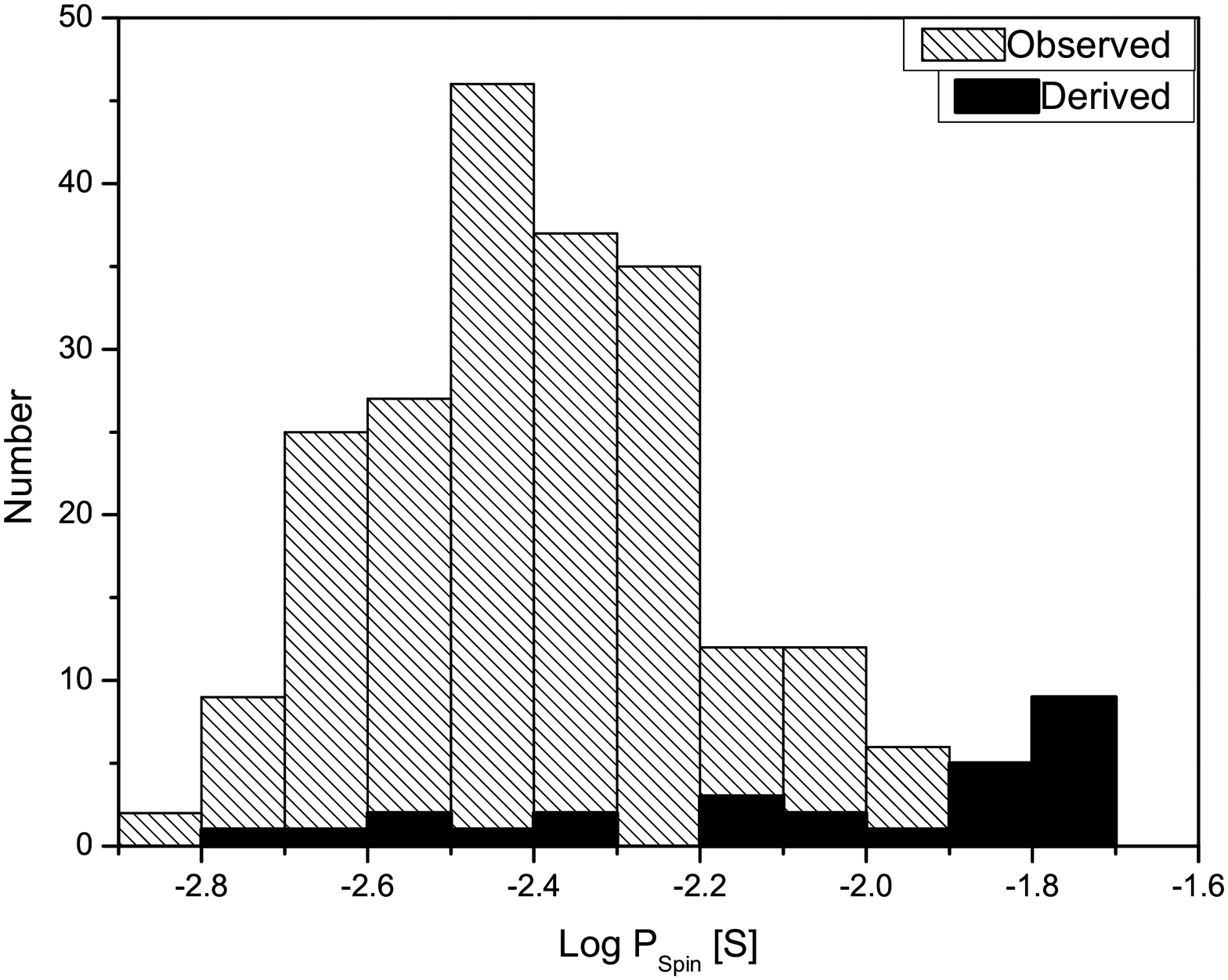}\hfill
\includegraphics[angle=0, width=4.5 cm]{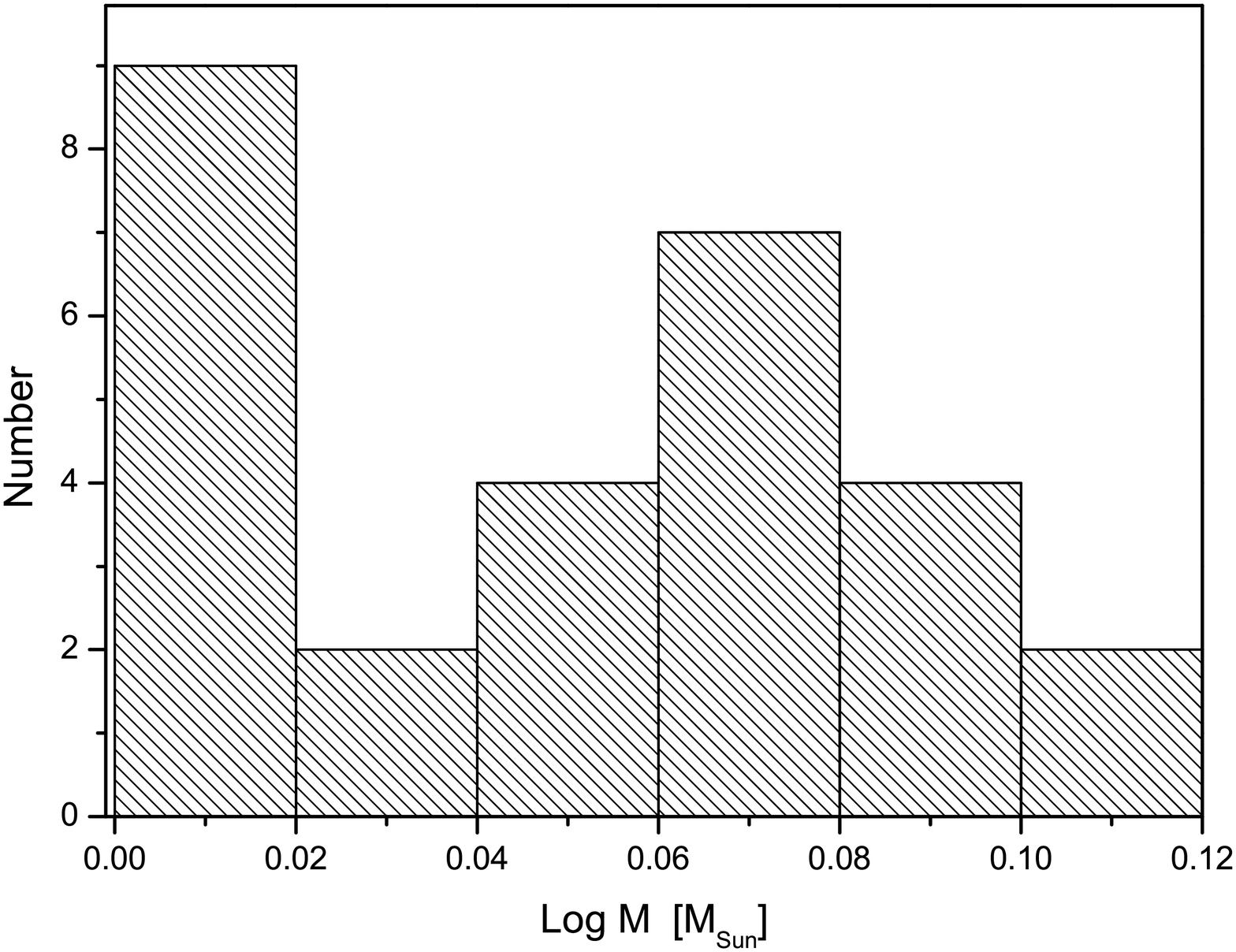}
\hspace*{\fill}
\end{center}
\caption{{\bf Left:} The distribution of MSPs in terms of spin-period (observed data from \cite{Man05}. 
{\bf Right:} Mass distribution of massive CVs \cite[(Ritter \& Kolb 2011)]{Rit11}.} \label{figure}
\end{figure*}

%


\section{Evolution of spin period}

We are able to determine the spin period of MSPs originating from
WDs. The process begins with simple Keplerian frequency. We assume
that the angular velocity of the NS is equal to the Keplerian
angular velocity, $v_{K}$, of the magnetosphere, at roughly the
Alfv$\acute{\textrm{e}}$n surface,
\begin{equation}
v_{K} \propto R_{NS}^{-3/2} \rightarrow P_{MSP} \sim R_{NS}^{3/2}
\end{equation}
from which we obtain the spin of the MSP, $P_{MSP}$, as a function of the WD spin, $P_{WD}$.
\begin{equation}
P_{MSP} \sim P_{WD, min} \left(\frac{R_{NS}}{R_{WD}} \right)^{3/2}
\end{equation}
where $P_{WD, min}$ is the minimum spin period of the standard WD \cite[(Warner 1995)]{War95}.
Assuming $P_{WD, min}$ $\sim 30$ s, $R_{NS} = 10$ km and
$R_{WD} = 1000 $ km, we thus obtain,
$  P_{MSP} \sim 1 ms. \nonumber $
Fig. 1 (left) shows the observed and derived spin period
distributions of
MSPs. 
As for the observed MSPs, they show
a relatively Gaussian distribution.
According to
these distributions, the ratio of MSPs originating from CVs is about
$\sim 10\%$. This result agrees with some theoretical predictions
such as those by \cite{War95} and \cite{War02}.

\section{Magnetic field}

To investigate the correlation between the magnetic fields of
MSPs with bottom fields of CVs (where the field is partially restructured due to accretion),
we follow the model of \cite{Zha09}. 
The magnetic
flux is assumed to be conserved. This corresponds to the magnetic fields produced by AIC. 
\begin{equation}
B_{NS} = B_{f,WD} \times \left(\frac{R_{WD}}{R_{NS}}\right)^2
\end{equation}
If we adopt
$R_{NS}$ = $15\times10^{5}$~cm
and $B_{f,WD}\sim10^3$G in CVs, the minimum value,
$ B_{NS}\sim10^{8.5-9}  G.
$





\section{Mass}

The sample of CVs whose masses we have considered is the set of
binary systems collected by \cite{Rit11}. Among them we have 26
massive CVs in the range ${\rm~1.0 - 1.3M_{\odot}}$. Fig. 1 (right)
shows the relatively Gaussian distribution, with mean at
${\rm~M_{CV}\sim 1.1M_\odot}$. A summary of the known properties of
these systems is given in Table 1 of \cite{Taa12}. Note that the AIC
process leads to a MSP with mass less than Chandrasekhar limit
\cite{Zha11}. This provides evidence for the AIC in massive CVs and
evolutionary hypotheses of MSP birthrate.

\section{Summary and Conclusions}


\begin{enumerate}
\item CVs would be invoked via the capability of producing a
significant portion of the MSPs via the AIC process,
a regime which may be unattainable by normal channels.

\item We find that the quantitative implications of our
calculations are that we estimate the expected $P$ in the observed of MSPs which could have originated from CVs to be $\sim 10\%$. 

%
%

\item We further find that the predictions of some parameters
after AIC process for the average levels are consistent with the
observed MSP population. Future work will consider
other quantities e.g. orbital period, eccentricity,
and mass ratio (q) using more data sets from SDSS.

\end{enumerate}

We are grateful for the discussion with Cole Miller. This research
has been supported by NBRPC (2009CB824800, 2012CB821800) and the
NSFC (10773017, 11173034).

%

{}


\begin{thebibliography}{}
%
%
%
%
%
%
%
%
%
%
%


\bibitem[Manchester et al.(2005)]{Man05}
{Manchester, R.N., Hobbs, G.B., \& Teoh, A., et al.} 2005 \textit{AJ} 129,
1993


\bibitem[Ritter \& Kolb (2011)]{Rit11}
{Ritter, H. \& Kolb, U.,} 2011, \textit{VizieR Online Data Catalog} 1,
2018



%


 \bibitem[Taani et al. (2012)]{Taa12}
{Taani, A., Zhang, C.M., \& Al-Wardat, M., et al.} 2012 \textit{ Ap\&SS}
340, 147



\bibitem[Warner (1995)]{War95}
{Warner, B.,} 1995, \textit{Cataclysmic Variable Stars. Cambridge
Astrophysics Series vol.} 28

\bibitem[Warner \& Woudt (2002)]{War02}
{ Warner, B. \& Woudt, P.A.,} 2002\textit{ The Physics of Cataclysmic
Variables and Related Objects. Astronomical Society of the Pacific
Conference Series} vol. 261

%


\bibitem[Zhang et al. (2009)]{Zha09}
{Zhang, C.M., Wickramasinghe, D.T., \& Ferrario, L., et al.} 2009
\textit{MNRAS} 397, 2208


\bibitem[Zhang et al. (2011)]{Zha11}
{Zhang, C.M., Wang, J., \& Zhao, Y. H., et al. } 2011 \textit{A\&A} 527,
83




\end{thebibliography}
\end{document}